 \documentclass[final,3p,times]{elsarticle}

\usepackage{amssymb}

\journal{Physica A}

\begin{document}

\begin{frontmatter}



\title{Disorder induced phase transition in kinetic models of opinion dynamics}

 \author[label1]{Soumyajyoti Biswas}
\address[label1]{Theoretical Condensed Matter Physics Division,
Saha Institute of Nuclear Physics, 1/AF Bidhannagar, Kolkata 700 064, India.}
\author[label2]{Arnab Chatterjee}
\address[label2]{Centre de Physique Th\'{e}orique (CNRS UMR 6207), Universit\'{e} de la M\'{e}diterran\'{e}e Aix Marseille II,
Luminy, 13288 Marseille cedex 9, France.}
\author[label3]{Parongama Sen}
\address[label3]{Department of Physics, University of Calcutta,
92 Acharya Prafulla Chandra Road, Kolkata 700009, India.}




\begin{abstract}
We propose a  model of continuous opinion dynamics, where mutual interactions can
be both positive and negative. Different types of distributions 
for the interactions, all characterized by a single parameter $p$ denoting  
the fraction of negative interactions, are considered. 
 Results from  exact calculation of a discrete version
and numerical simulations  of the continuous version of the model
indicate  the existence of a universal continuous phase transition at $p=p_c$ 
below which a consensus is reached. 
Although the order-disorder transition is analogous to a ferromagnetic-paramagnetic phase transition with comparable critical exponents,
the model is characterized by some  distinctive features relevant to a social system.
\end{abstract}


\end{frontmatter}


\section{Introduction}\label{sec:1}

\noindent Quantitative understanding of individual and social  dynamics  
has been explored on a large scale \cite{ESTP,Stauffer:2009,barabasi,vicsek,redner,Castellano:RMP,cohen,Galam:1982} in recent times.
Social systems offer some of the richest complex dynamical systems, 
which can be studied using the standard tools of statistical 
physics. With the availability of  data sets and records on the increase, microscopic models mimicking these systems  
help in understanding their underlying dynamics. On the other hand, some of these models 
exhibit novel critical behavior, enriching the theoretical aspect of these studies.

Mathematical formulations of such social behavior have helped us to 
understand how global consensus (i.e., agreement of opinions)  emerges out of individual opinions
\cite{Holme:2003,Fortunato:2005b,Krapivsky:2003,Kuperman:2002,Liggett:1999,Sznajd:2000,Galam:2008,galam3,Deffuant:2000,Hegselman:2002,Fortunato:2005,Toscani:2006,galam1,Sen:opin,Lallouache:2010,Sen:2010,bcc}.
Opinions are usually modeled as variables, discrete or continuous, and
are subject to spontaneous changes as well as changes due to binary interactions, global feedback and even external factors.
Apart from the dynamics, the interest in these studies also lies in the distinct steady state properties:
a phase characterized by individuals with widely different opinions and another phase
with a major fraction of individuals with similar opinions. Often the phase transitions are driven by appropriate parameters of the model.

In this paper we study a model of opinion dynamics by considering 
two-agent interactions. 
Continuous opinion dynamics has been studied for a long time \cite{stone,degroot,seneta}, with the models
designed in such  a way that eventually the opinions cluster around 
one (consensus), two (polarization) or many (fragmentation) values.
The average opinion or macroscopic behavior have been emphasized 
only in some recent works \cite{Lallouache:2010, Sen:2010}, where a phase transition from ordered
to disordered phase has also been reported.
  However,
in contrast to these models, we obtain here an ordered phase where even in the presence
of a dominant opinion (symmetry broken phase), opposing opinions survive and a disordered phase where all opinion values
coexist without any preference to any value (symmetric phase). Thus we present this in the general
context of an order-disorder transition similar to that of the Ising and related models. 
We also compare our results with earlier works where a mean-field phase transition was observed
in presence of contrarians in the society \cite{galam1}.

The paper is organized as follows: in Section 2 we introduce the model. Then in Section 3
the main results are presented along with the calculations and numerical simulations.
In Section 4 we extend the model to include bond dilution and present the phase diagram. Finally we discuss
our results in Section 5.
  
\section{The Model}\label{sec:2}

We propose a new model for emergence of consensus.
Let $o_i(t)$  be the opinion of an individual $i$ at time $t$.
In a system of $N$ individuals (referred to as the `society' hereafter),
opinions change out of pair-wise interactions:
\begin{eqnarray}
\label{eq:model}
 o_i(t+1) &=& o_i(t) + \mu_{ij} o_j(t).
\end{eqnarray}
One considers a similar equation for $o_j(t+1)$.
The choice of pairs $\left\{i,j\right\}$ is unrestricted, and hence
our model is defined on a fully connected graph, or in other words, of infinite range.
Note that this is simply a pair-wise interaction and we imply no sum over the index $j$.
Here $\mu_{ij}$
are real, 
and it is like an interaction  parameter representing the   influence of 
the individual with whom interaction is taking place. 
The opinions are bounded, i.e., $-1 \le o_i(t) \le 1$. This bound, along with Eq.~(\ref{eq:model}) defines the dynamics of the model.
If, by following Eq.~(\ref{eq:model}) the opinion value of an agent becomes higher (lower) than $+1$ ($-1$), then it is 
made equal to $+1$ ($-1$) to preserve this bound.
The ordering in the system is measured by
the quantity $O=  |\sum_i o_i |/N$, the average opinion, which is the order parameter for the system. 

The present model is similar in form to a  class of simple models  proposed 
recently~\cite{Lallouache:2010,Sen:2010,bcc,sb},
apparently inspired by  the kinetic models of wealth exchange~\cite{CC-CCM,Chatterjee:2009}.
A spontaneous symmetry breaking was observed in such models: in the symmetry
broken phase, the average opinion is nonzero while in the symmetric phase,
the opinions of all individuals are identically zero indicating a `neutral state'. 
The parameters representing conviction (self interaction) and influence (mutual interaction)
  in these models were considered either  uniform (a scalar) or  in the generalized case   
different for each individual, i.e, given by the components of a vector.
In addition to this there is an added feature of the randomness in 
the influence term which in effect controls the sharpness of the phase transitions
in these models.

In our proposed model, the  conviction parameter or self interaction parameter  is set 
equal to unity  so that in absence of interactions, opinions remain frozen.  
In such a situation, it has been observed previously that any
interaction, however small, leads to
a highly unrealistic state of all individuals having extreme identical opinions (either
 $o_i=1$ $\forall$ $i$  or $o_i = -1$ $\forall$ $i$)~\cite{Sen:2010} 
when the interactions take up  {\it positive values only}.
This suggests that one should generalize the
interactions to include both positive and negative values. This is realistic also in the
sense that it reflects the fact that in a social interaction of two individuals, there may be
either agreement or disagreement of opinions.   
We therefore consider not only a distribution of the values of $\mu_{ij}$ (to maintain the stochastic nature
of the interactions) but also allow $\mu_{ij}$ to have negative values.
We define a parameter $p$ as the fraction of values of $\mu_{ij}$ which are negative, 
which, we will show later, leads to characteristic
ordered and disordered states as in reality.

The fact that we allow random positive and negative values for the interactions may 
suggest that the model is analogous to a  dynamic spin glass model~\cite{SpinGlass,Nishimori},
as in the latter, one can consider a dynamic equation for the spins which formally
resembles Eq.~(\ref{eq:model}).
 However, the two dynamic models are not equivalent with the following differences:  
(i) the interactions in the opinion dynamics models are never considered simultaneously and 
thus the question of competition leading to the possibility of  frustration does not arise, and (ii) there is also no energy function to  minimize, (iii) the symmetry
$p\to1-p$ does not exist in our model, which is naturally present for spin-glass. 
We will get back to the comparison of the two models in the context of phase transition later 
in this paper.

The effect of negative interactions was considered previously in a different opinion dynamics model
under the name Galam contrarian \cite{galam1}. The discrete, binary opinion model followed a deterministic 
evolution rule for a group
of three or more individuals. It was shown that depending on the concentration of the contrarians, the
system will either reach an ordered state, where there one of the opinions will have majority, or a disordered
state, where no clear majority is observed. The critical behavior of the model is similar to the one we present here
at least in the fully connected graph. However, our model considers continuous opinion values. Also,  the Galam
contrarians always take the opinion opposite to that of the majority. However, in our case we also consider
the present state of opinion of the agents and accordingly even the discrete version of our model has three states.
A two-state discrete version of this model will not show any ordered state. 

\section{Results}

Unless otherwise mentioned, we keep
$\mu_{ij}$ values within the interval $[-1,1]$ for simplicity.   
In principle, several forms can
be considered for $\mu_{ij}$ (annealed, quenched, symmetric, non-symmetric
etc.). Further, there can be several distribution properties
for $\mu_{ij}$ in the interval $[-1,1]$ (discrete, piecewise uniform and continuous distributions). 
Unless otherwise stated, in our study, we would discuss the case when  $\mu_{ij}$
are \textit{annealed}, i.e., they change with time. In other words,
at each  pairwise interaction, the  value of $\mu_{ij}$ is randomly chosen  respecting the fact
that it is negative with  probability $p$.
For this case, the issue of symmetry does not arise.
We consider distributions for both continuous and discrete $\mu_{ij}$.

In all the above cases, we find a symmetry breaking transition.
Below a particular value $p_c$ of the parameter $p$, the system orders (i.e., the order parameter $O$ has a finite non-zero value), while the disordered phase
(where $O$=0) exists for higher values of $p$.
Since this phase transition is very much like the thermally driven ferromagnetic-paramagnetic transition in magnetic systems, 
we have considered the scaling of the analogous static quantities, which are:\\
\noindent
(i) the average order parameter $\langle O\rangle$, $\langle \ldots \rangle$ denoting average over configurations,\\
\noindent
(ii) the fourth order Binder cumulant $U = 1-\frac{\langle O^4 \rangle}{3\langle O^2 \rangle^2}$,\\
\noindent
(iii)  a quantity analogous to susceptibility per spin, which we  write as  $V=N[\langle O^2 \rangle - \langle O \rangle^2$].

We also calculate
(iv) the condensate fraction $f_c=f_1 + f_{-1}$,
where $f_1$ and $f_{-1}$ denote the fraction of population with opinion $1$ and $-1$ respectively.
$f_c$ is exclusive to this class of  opinion dynamics models and is  
expected to show scaling behavior near the critical point~\cite{Lallouache:2010,Sen:2010}.

\begin{figure}[t]
\centering \includegraphics[width=8.7cm]{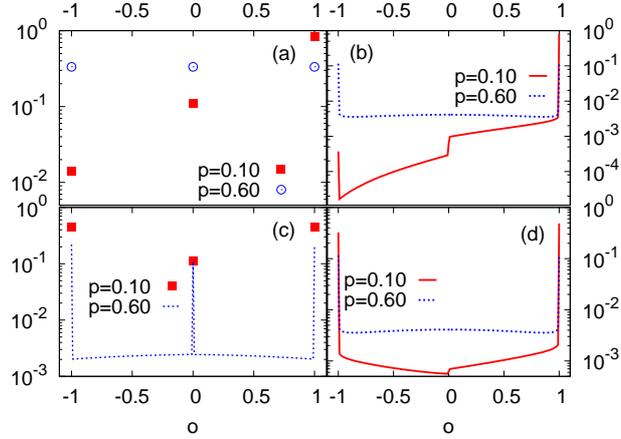}
\caption{
Probability distribution of opinions for polarized (all $+1$) 
initial condition with (a) discrete $\mu_{ij}$ ($p_c=1/4$) and
(b) continuous $\mu_{ij}$ ($p_c\approx 0.34$); same for random
(uniform between [$-1:+1$]) initial condition with (c) discrete $\mu_{ij}$ ($p_c=1/4$) and
(d) continuous $\mu_{ij}$ ($p_c\approx 0.34$).
All data are for $N=4096$.}
\label{fig:opd}
\end{figure}

Before a more detailed characterization of this transition, we find it useful to describe the behavior of the
probability distribution of the order parameter first. This
distribution  itself shows the signature of a phase transition. We consider both the cases with  discrete ($\pm 1$)  and 
continuous (within the bound $ -1 \le \mu_{ij} \le +1 $) values for $\mu_{ij}$.
For each of the above case we consider both polarized and random initial conditions. We show
the cases where the initial condition is fully polarized (all
$+1$) and $\mu_{ij}$ are discrete (Fig.~\ref{fig:opd}(a)) and continuous (Fig.~\ref{fig:opd}(b)). 
We consider two $p$
values, one below and one above the critical point. As can
be seen from the figures, for $p = 0.6 > p_c$ the distributions
are symmetric in both the cases, which manifestly imply
disorder, while the ordered phase is asymmetric with a
bias towards $+1$ due to the initial condition.

Now consider the case when the initial condition is random instead of being polarized as above. Then of course
in a particular ordered state either majority positive or
majority negative opinion can occur. Hence the order parameter distributions 
(after averages over many configurations) become symmetric (Figs~\ref{fig:opd}(c),(d)) even in the 
ordered phase for both discrete and continuous $\mu_{ij}$ values.
An interesting point, however, is the depletion of 
population in the intermediate (between the extremist with 
opinions $\pm 1$) opinion values in an ¡¥ordered¡¦ society (Fig.\ref{fig:opd}(c)).
This is even more pronounced when we make $\mu_{ij}$ values
discrete ($\pm 1$). In this case, the distribution becomes 
discretized almost simultaneously with the symmetry 
breaking transition such that the intermediate opinion values
do not exist at all in the ordered state (Fig.\ref{fig:opd}(d)). 
Therefore, we find that in the ordered state with discrete $\mu_{ij}$,
the society becomes highly clustered in the sense that 
continuous variation of opinion is no longer a possibility. This
of course is commensurate with the continuous $\mu_{ij}$ case 
described above, where possibility of ordering leaves the 
society highly clustered as is generally seen before and after
decisive elections.

It is important to note here that although the distribution of average opinion is sensitive to the initial condition, but
the order parameter itself is not as the latter is obtained by taking the absolute
value of the average  opinion in a single configuration and then taking a further average over all configurations in a numerical study.
Moreover, the critical behavior is also unaffected by the changes in initial condition.

Our major interest lies in identifying the critical behavior in these models.
Assuming that there exists a steady state for these models (which are numerically observed), 
one can derive exact expressions for the steady state probabilities $f_1$, $f_0$ and $f_{-1}$
 in the annealed \textit{discrete} case where we assume the initial condition to be such 
that the agents have $o_i(t=0) \in \{-1,0,+1 \}$. The scaling behavior of 
the order parameter  and $f_c$ can be exactly obtained in this method.
To do that, we consider the probabilities that an agent's opinion gets decreased (for initial opinion $+1$ or $0$)
or increased (for initial opinion $-1$ or $0$) or remains constant when two agents interact.    
For example, let us consider the change for the  agent A, interacting with a second agent B 
(one can consider updated values of both but it does not matter): When both have opinion $+1$ (probability of occurrence $f_1^2$), 
 A's opinion decreases with a probability $p$, giving the joint occurrence probability of the event as $pf_1^2$. 
On the other hand, when B has opinion $-1$,
 the similar event has the  probability $(1-p)f_1f_{-1}$.  
One can similarly find out the probabilities of decrease and increase  in all possible cases. 
The exact expression for the
net increase probability is 
$p f_{-1}^2 + (1-p) f_1 f_0 +p f_0 f_{-1} + (1-p) f_1 f_{-1}$
 and that for the net decrease probability  
is $p f_1^2 + p f_1 f_0 + (1-p) f_0 f_{-1} + (1-p) f_1 f_{-1}$.
In the steady state these two should be equal, i.e.,
\begin{equation}
p f_1^2 + p f_1 f_0 + (1-p) f_0 f_{-1}
         = p f_{-1}^2 + (1-p) f_1 f_0 + p f_0 f_{-1},
\end{equation}   
which simplifies to 
\begin{equation}
(2f_1+f_0-1)\left[p-f_0(1-p)\right] = 0 .
\end{equation}
This means, either $2f_1+f_0=1$, i.e., $f_1=(1-f_0)/2=f_{-1}$ which implies a  disordered phase, or
\begin{equation}
f_0=\frac{p}{1-p}.
\label{eq:pf0}
\end{equation}
Next we show  that at criticality all three fractions would become equal such that $p_c=1/4$. Let us  
 take the solution in the disordered phase
where $f_1=f_{-1}$.
We  consider the processes contributing to the in/out flux for $f_0$. 
We enumerate all possibilities as before and get the following:
flux into $f_0$ is $2[(1-f_0)/2]^2$ and flux out of $f_0$ is $f_0(1-f_0)$.
So, in the steady state
\begin{equation}
\left( \frac{1-f_0}{2} \right)^2=\frac{f_0(1-f_0)}{2}.
\end{equation}
Hence, either $f_0=1$, which can  be ignored by considering the steady states of the
other two fractions, or $f_0=1/3$.
Now, just at the critical point this 
solution will begin to be valid. 
Hence, at critical point all three fractions are $1/3$, leading to $p_c=1/4$. 

\begin{figure}[t]
\centering \includegraphics[width=8.7cm]{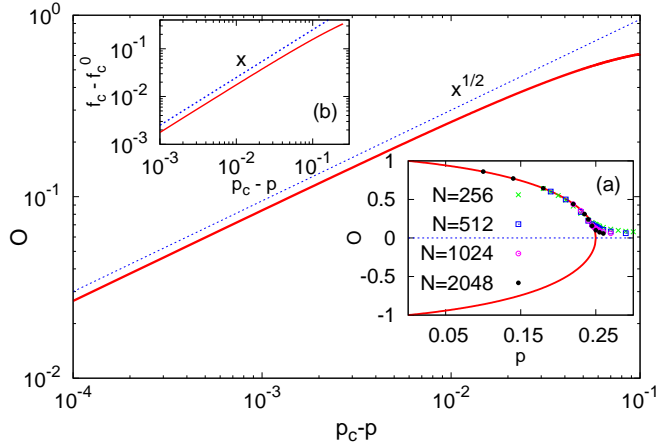}
   \caption{Discrete $\mu_{ij}$:
power law behavior of the order parameter $O$ near the critical point $p_c$ (Eq.~(\ref{mfeq9}))
showing $\beta=1/2$. The dotted line is $x^{1/2}$, a visual guide.
Inset: (a) Phase diagram. The points represent simulation results. They are in better agreement with the analytical results as the system size is increased from 256 to 2048.
(The lower half of the phase diagram follows from symmetry.) 
(b) Linear scaling of $f_c-f_c^0$. The dotted line is $x^1$.}
\label{fig:mf}
\end{figure}
In the ordered phase, the order parameter $O$ is given by $|f_1 - f_{-1}|$.
To evaluate $f_1 $ and $f_{-1}$, we calculate the flux in and out of $f_1$. 
 Flux out of $f_1$ is $pf_1^2+(1-p)f_1[1-(f_1+f_0)]$ and 
 flux into $f_1$ is $(1-p)f_0f_1+ pf_0 [1-(f_1+f_0)]$. Hence at steady state,  
$f_1$ is given by 
\begin{equation}
f_1=\frac{1-3p+2p^2\pm \sqrt{1-6p+9p^2-4p^3}}{2(1-2p+p^2)},
\end{equation}
where we have used Eq.~(\ref{eq:pf0}). 
Hence the order parameter is given by
\begin{equation}
O=\frac{1-3p+2p^2\pm \sqrt{1-6p+9p^2-4p^3}}{(1-p)^2}+\frac{2p-1}{1-p}.
\end{equation}
The variation of the order parameter with $p$ is shown in Fig.~\ref{fig:mf}. 
It shows the expected behavior, i.e.,  it  vanishes at $p=p_c$. 
Rewriting the above equation in terms of $x=p_c-p$,   algebraic simplifications give
\begin{equation}
O=\frac{3/8+2x+2x^2\pm \sqrt{9x/4-3x^2}}{9/16+3x/2+x^2}-\frac{2x+1/2}{3/4+x}.
\label{mfeq9}
\end{equation}
As $x\to 0$, $O \sim \sqrt{x}$, 
implying the critical exponent for $O$ is $\beta=1/2$. 
This also agrees well with the power law fit of the order parameter expression 
(Eq.~(\ref{mfeq9})) near the critical point (Fig.~\ref{fig:mf}).
Calculation for $f_c$ on the other hand shows that it has a constant  
value $f{_c}^0 =  2/3$ beyond $p_c$ and 
for $p< p_c$, 
$f_c -f{_c}^0  \sim x^{1}$, i.e., vanishes linearly at $p_c$, 
which also perfectly agrees with numerical simulations.
Numerical simulations for the model with continuous $\mu_{ij}$ 
yields similar behavior with $f{_c}^0 \simeq 0.22$ since opinion values other
than $\pm 1$ and $0$ exist.

\begin{figure}[t]
\includegraphics[width=7.5cm]{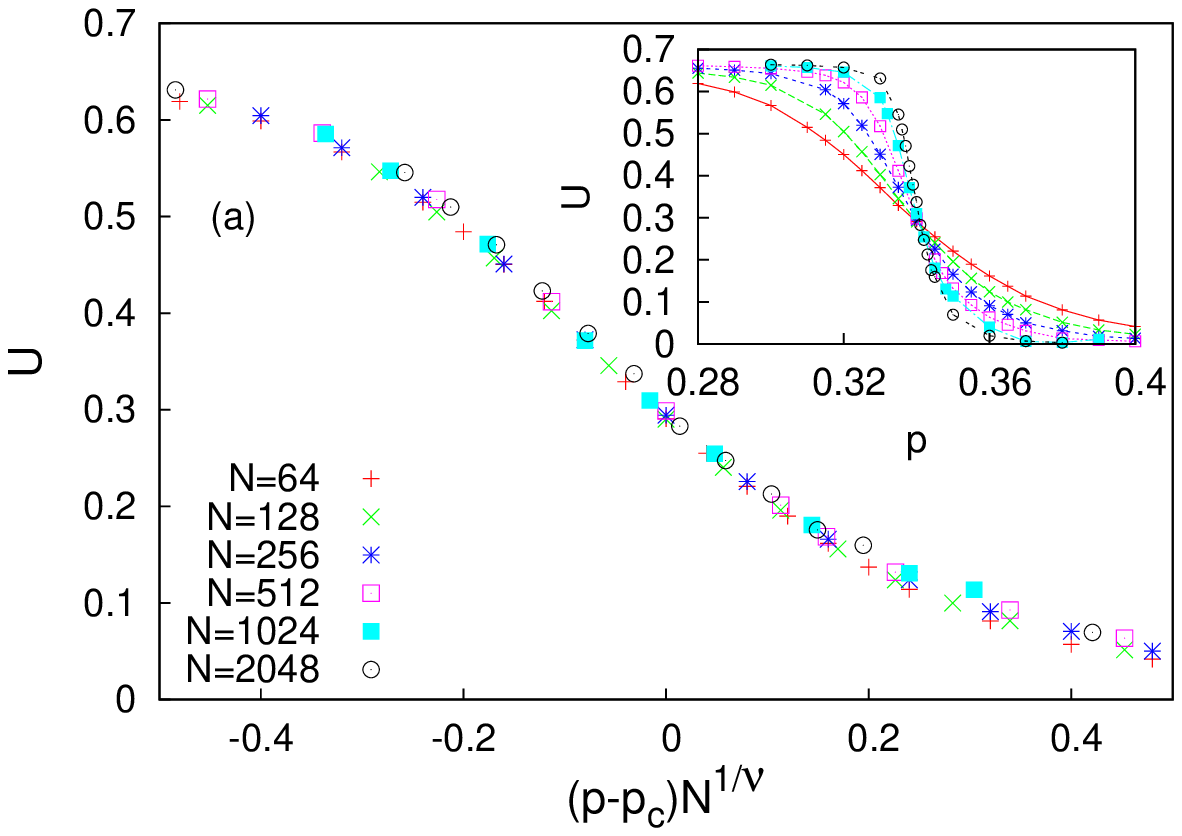}
\includegraphics[width=7.5cm]{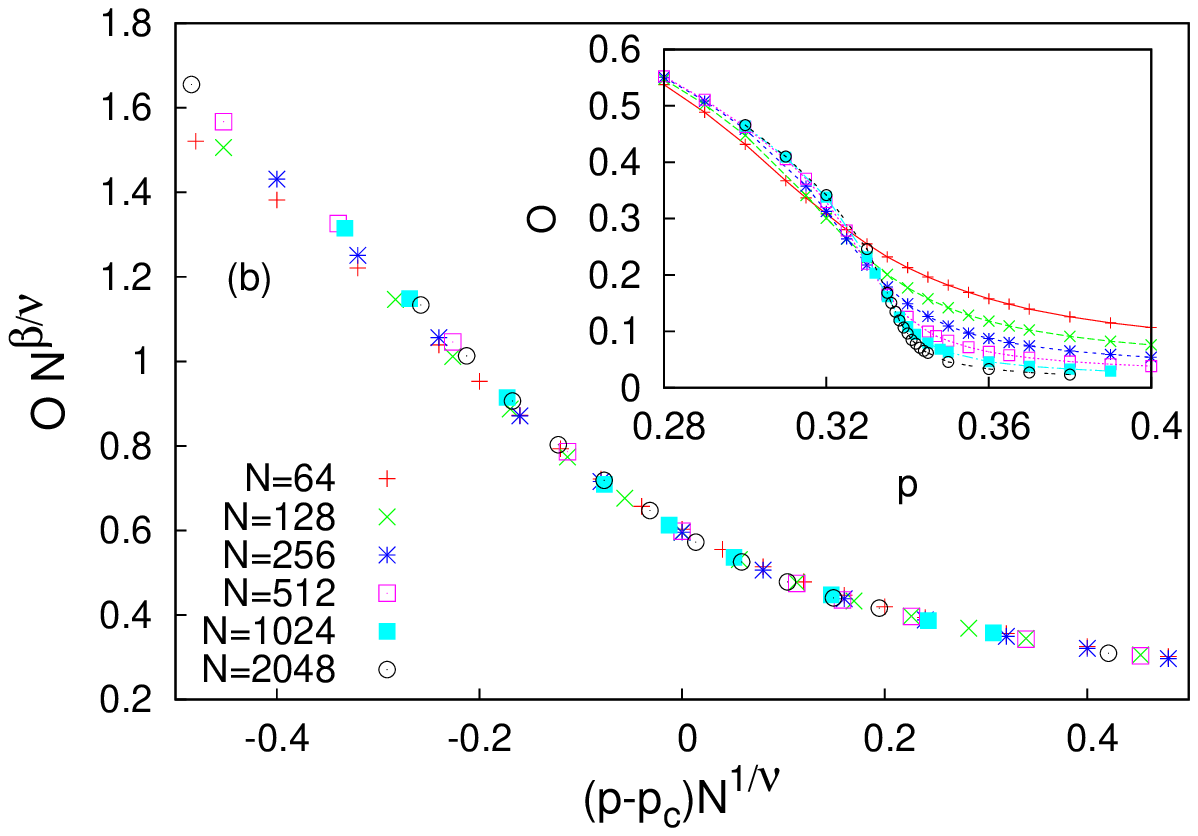}
\includegraphics[width=7.5cm]{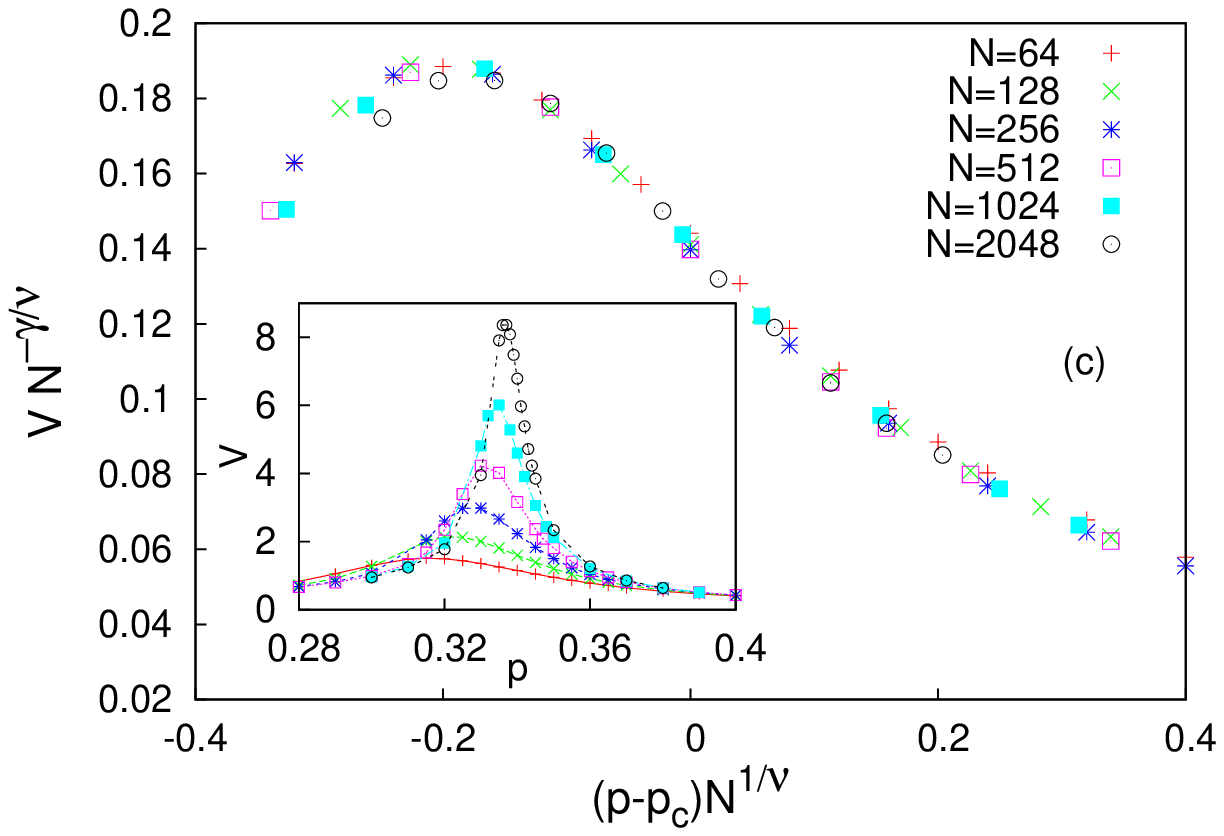}
\caption{Data for continuous, annealed $\mu_{ij}$ model, showing
(a) finite size scaling of the Binder cumulant $U$ for different system sizes $N$;
 the critical point is $p_c = 0.3404\pm0.0002$,  and the best data collapse is for $\nu=2.00\pm 0.01$.
Inset: Variation of $U$ with $p$ for different system sizes; 
(b) finite size scaling of order parameter $O$ for different $N$;
 best data collapse is for 
$\beta=0.50 \pm 0.01$.
Inset: Variation of the order parameter $O$ with $p$ for different system sizes;
(c) finite size scaling of $V$  for different $N$;
  best data collapse is for 
$\gamma=1.00 \pm 0.05$.
Inset: Variation of $V$ with $p$ for different $N$. Number of averages for
different system sizes are 3000 for $N=256$, 1800 for $N=512$, 1000 for $N=1024$ and $400$ for $N=2048$.}
\label{fig:contannU}
\end{figure}
\begin{figure}[t]
\includegraphics[width=7.5cm]{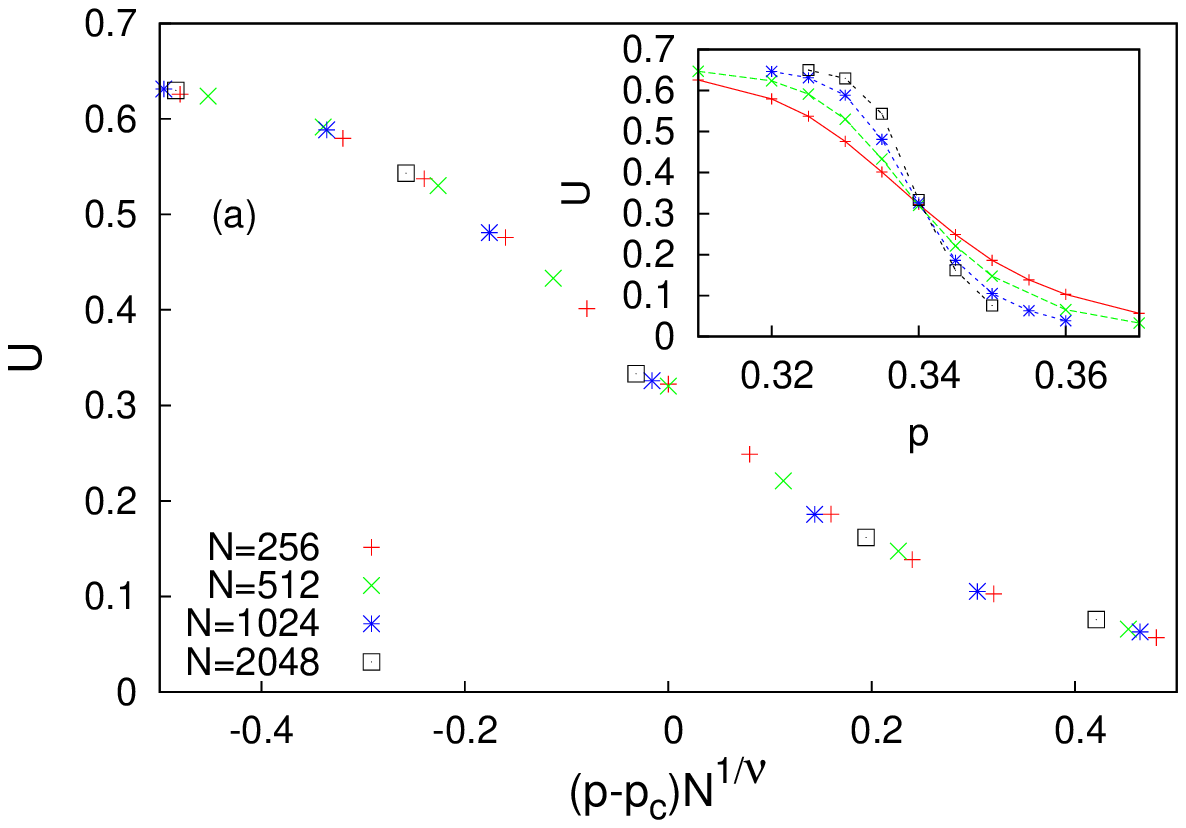}
\includegraphics[width=7.5cm]{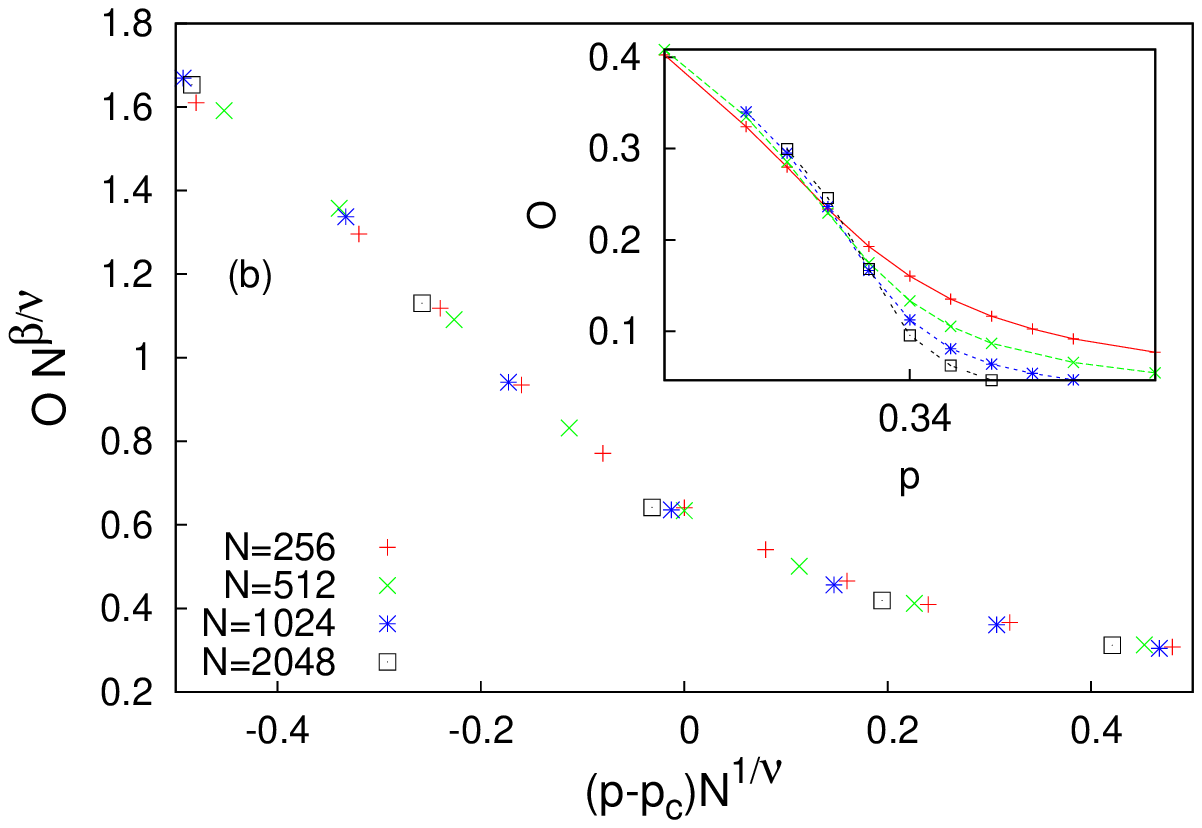}
\includegraphics[width=7.5cm]{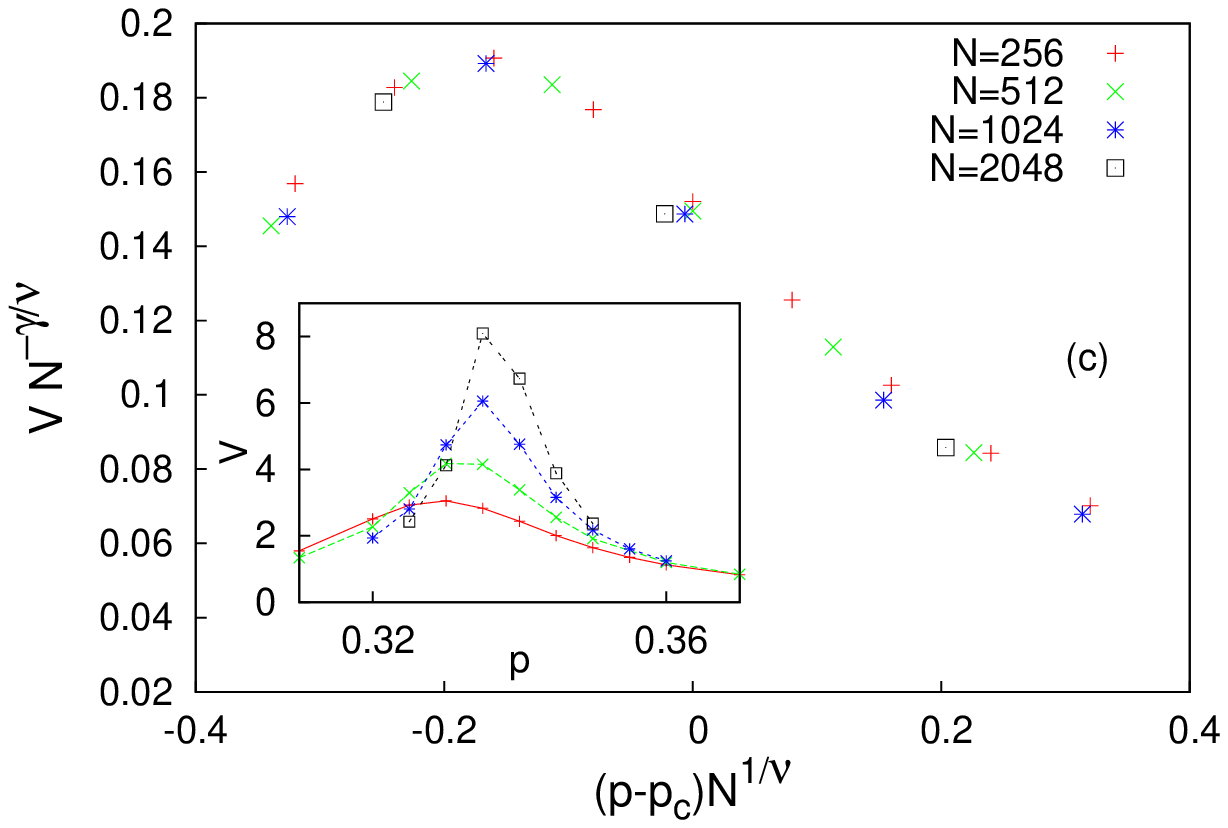}
\caption{Data for continuous, quenched $\mu_{ij}$ model, showing
(a) finite size scaling of the Binder cumulant $U$ for different system sizes $N$;
 the critical point is $p_c = 0.34\pm0.01$,  and the best data collapse is for $\nu=2.00\pm 0.01$.
Inset: Variation of $U$ with $p$ for different system sizes; 
(b) finite size scaling of order parameter $O$ for different $N$;
 best data collapse is for 
$\beta=0.50 \pm 0.01$.
Inset: Variation of the order parameter $O$ with $p$ for different system sizes;
(c) finite size scaling of $V$  for different $N$;
  best data collapse is for 
$\gamma=1.00 \pm 0.05$.
Inset: Variation of $V$ with $p$ for different $N$.Number of averages for
different system sizes are 3000 for $N=256$, 1800 for $N=512$, 1000 for $N=1024$ and $400$ for $N=2048$.}
\label{fig:contqueU}
\end{figure}
Using the same kind of argument as above, one can   show that there will be 
no phase transition when $\mu_{ij}=\pm \mu_0$ with $\mu_0\ge 2$.
It is obvious that in this case the opinions can have only two values $+1$ and $-1$. 
Let $f_1$ be the fraction of agents having opinion +1.
Once again considering net decreases and increases of opinions and equating 
them in the steady state, we get
\begin{equation}
p(1-f_1)^2=pf_1^2.
\label{diso}
\end{equation}
For any non-zero $p$ the solution of this equation is $f_1=1/2$ thereby giving complete 
disorder. Hence, in this condition there cannot be an ordered phase for any finite $p$.

We performed Monte Carlo simulation  for
different system sizes ($N=64,256,512,1024,2048$) to estimate $p_c$ and all the
relevant exponents for discrete as well as continuous $\mu_{ij}$'s (see Fig.~\ref{fig:contannU} 
where the data for the continuous distribution of $\mu_{ij}$ are presented). 
A Monte Carlo step is the simultaneous update of $N$ agent's opinion values.
For each simulation point, sufficient relaxation time was given (depending on system size),
such that the measurable quantity reached a steady-state value. Then the ensemble 
average of those steady state values were taken (number of ensemble again depends 
on system size, see figure captions for details).

We estimated the critical point $p_c$ from the crossing of the Binder cumulants
for different system sizes~\cite{Binder:1981}. Our estimate is 
$p_c \simeq 0.249 \pm 0.001$ for the discrete case  which is consistent
with the analytical value of $1/4$ derived earlier. 
The critical Binder cumulant is $U^*=0.30\pm0.01$.
For the continuous case, $p_c = 0.3404 \pm 0.0002$, and $U^*=0.284\pm0.004$.  
We find excellent finite size collapse for all cases.
We estimate the correlation length exponent $\nu = 2.00\pm 0.01$, the order parameter exponent
$\beta= 0.50\pm 0.01$  and the fluctuation exponent $\gamma = 1.00\pm 0.05$.

One can consider several other distributions with a similar parameter $p$ 
to test the universality of the
phase transition: e.g., we take $\mu_{ij} = 1$ with probability  
$(1-p)$ and  $\mu_{ij} = -1/2$ with probability $p$. The 
transition point shifts but the exponents remain same. 

We also conducted detailed simulations for the case  where $\mu_{ij}$ are quenched variables, i.e., they do not change with time. 
For the fully connected graph considered here, numerical results for continuous $\mu_{ij}$'s (see Fig.~\ref{fig:contqueU}) indicate that critical behavior is the
same as that of the annealed case mentioned above.

\section{Model with bond dilution}
Here we briefly consider the case when all the interactions 
between the agents are not allowed, i.e., agents are selective in their interactions.
In the case of continuous distribution of $\mu_{ij}$, it does not make  sense
to parametrize  the probability of $\mu_{ij}  = 0$.   It is therefore useful
to consider the discrete distribution only  for the dilute case by 
 considering $\mu_{ij}=0 $ with probability $q$ along with  
 $\mu_{ij}=-1$ with probability $p$ as before and $\mu_{ij}=1$ otherwise.
Thus, $q$ is naturally the `bond dilution' parameter.

\begin{figure}[hhh]
\centering \includegraphics[width=8.0cm]{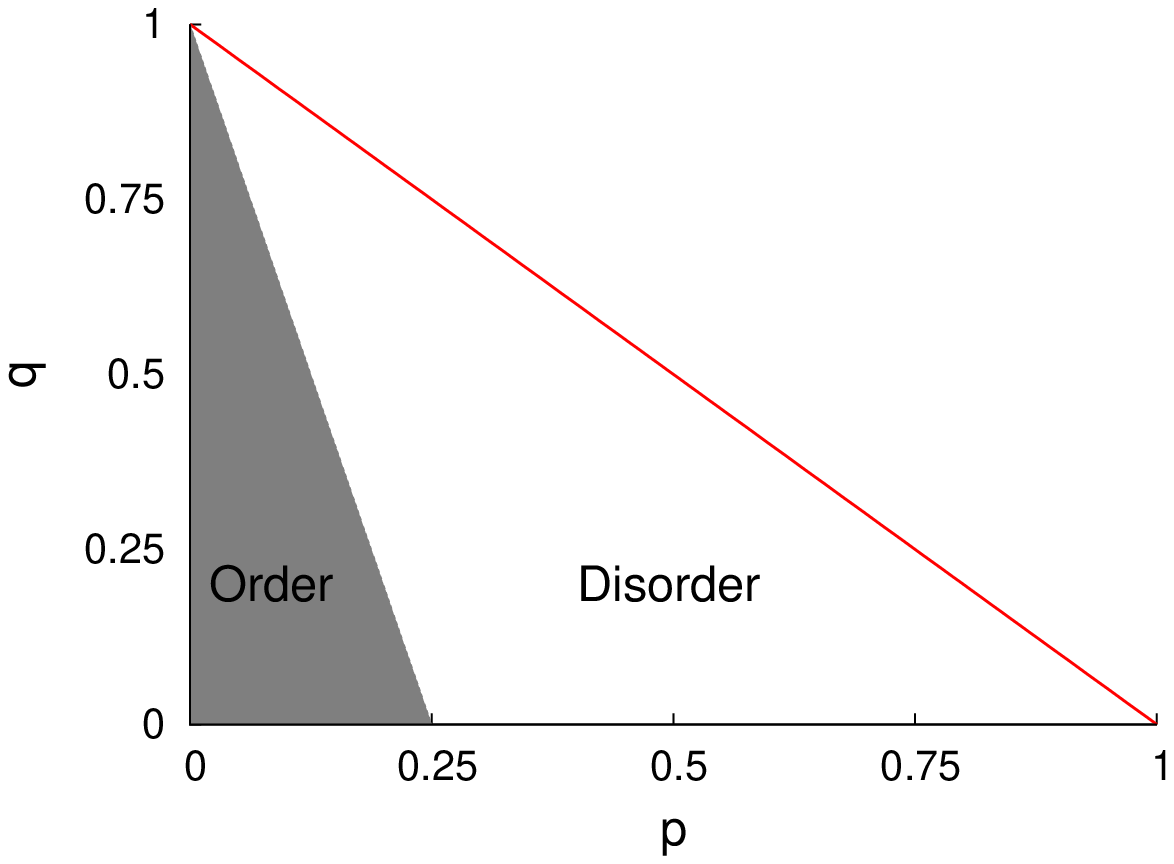}
\caption{Phase diagram for the dilute model.}
\label{fig:dilute}
\end{figure}

Considering the balance of the increases and decreases 
in the order parameter in the steady state, one gets
\begin{eqnarray}
f_1^2p &+& f_0f_1p+f_0f_{-1}\left[ 1-(p+q) \right] \nonumber \\
&=&f_{-1}^2+f_0f_1\left[ 1-(p+q)\right] +f_0f_{-1}p.
\end{eqnarray}
This gives, either $f_1=f_{-1}$, i.e., disorder, or in the ordered state
\begin{equation}
f_0=\frac{p}{1-p-q}.
\end{equation}
Now, in and out fluxes of $f_0$ gives
\begin{eqnarray}
 f_1^2p &+& 2f_1f_{-1}\left[1-(p+q)\right]+f_{-1}^2p= f_0f_1\left[1-(p+q)\right]  \nonumber \\
&+& f_0f_1p +f_0f_{-1}\left[1-(p+q)\right] +f_0f_{-1}p. 
\end{eqnarray}
In the disordered state, the only feasible solution is $f_0=\frac{1}{3}$. 
As before, assuming the continuity of $f_0$ across the transition point, we get
the phase boundary equation as
\begin{equation}
p_c=\frac{1}{4}\left(1-q_c \right),
\end{equation}
where $q_c$ is the critical value of $q$ on the phase boundary.
This phase boundary is shown in Fig.~\ref{fig:dilute}. 
Clearly $q=0$ limit is the unrestricted case studied
in the previous section. The fact that the ordered phase extends to $q=1$ is
for $p=0$ is consistent with the fact that for an infinite dimensional
lattice, percolation transition takes place at $q=1$.
                                            The universality of the model along the phase boundary needs to be done. One can expect it to be same as before.
\section{Discussions}\label{sec:3}
We present a simplified model for opinion formation in a society of highly 
connected individuals. With the dynamical rule defined by Eq.~(\ref{eq:model}), 
an agent modifies his/her opinion under the influence of another. This evolution rule
corresponds to cases where opinions are modified following discussions/debates with 
another individual. The key feature of our model is the inclusion of negative interactions with
probability $p$.
We study the steady-state collective behavior of this model. While modeling the opinion formation,
the effects of `self-conviction' and `external-pressure' have been considered before (see e.g., \cite{Lewenstein:1992}).

Here we concentrate upon the study of `consensus formation' as a dynamical critical phenomenon.
Eq.~(\ref{eq:model}) of course does not include all social complexities involved 
in such interactions, however in this simple form it manifests some intriguing features.

With the introduction of a single parameter $p$, defined in a simple manner, the proposed   model
 shows the existence of a universal phase transition and also some additional desirable features representing a real society.  
The parameter $p$ plays a role of the disordering field
 (similar to temperature
in thermally driven phase transitions). 
Beyond a certain value of the fraction of negative interaction, a phase transition 
from an `ordered' state (with  most of the individuals having opinions of same sign) 
to a `disordered' state (where the opinions can have  different signs and add up to zero) occurs
(see Eq.~(\ref{mfeq9})). 

The highlight of our model (with discrete values of $\mu_{ij}$) lies in the unique selection of
an ordered state with discrete opinion values ($\pm 1$ and $0$) while
in the disordered state, all opinion values in $[-1,+1]$ coexist (Fig.~\ref{fig:opd}(c)).
The disordered state  is one with a lot of disagreement, hence
all types of opinions co-exist in the society. But as it starts to get ordered (below $p_c$),
polarization occurs, and marginal opinions cease to exist, resembling the ordering
in a multi-party election scenario.
This unique selection of the ordered state
is in fact independent of the initial conditions which
adds to the richness of the model.
This of course also happens  in the bounded confidence models where
opinions, originally varying continuously, ultimately cluster around a few 
values typically. However, as already mentioned, in these models,
the dynamics is designed to achieve this, whereas in our model, it happens 
naturally 
 without any imposed restrictions.

The phase transition in our  model presents a case of a  classical continuous phase transition
with simple exponent values and showing finite size scaling behavior.
 To model `social temperature'
which essentially destroys consensus rather than forming it, we have kept negative interactions
among the agents. This leads to the desirable feature that  even  in the `ordered'
phase,  opinions of both signs can coexist and that the disordered state is also not a 
`neutral state'.
In contrast, in the models with only positive interactions~\cite{Lallouache:2010,Sen:2010,bcc},
 where  each `scattering' leaves the two agents closer
to each others' opinion,  finite size behavior was absent and the order of phase transition difficult to identify. 
Thus, we believe our model is closer to reality and also in
terms of critical behavior presents significant modifications.

Another important  point  to be mentioned is that one could change the distribution of the interactions in a way
such that no `neutral' individual remains in the society (opinion values are $\pm 1$).
It is intriguing and to some extent counter intuitive that such processes
lead to no consensus in our model (see Eq.~(\ref{mfeq9})).
When opinions can take values equal to $\pm 1$ only, a comparison with a Ising spin model is
bound to arise. The model with which one should compare is  
the fully connected Ising model with random bimodal distribution of the interactions 
($-J$  with  probability  $p$ and $+J$ with probability $(1-p)$).
However, in that model there is indeed a phase transition occurring
at $p=1/2$ \cite{Nishimori} from   a ferromagnetic to a spin glass phase. 
On the other hand, we get a transition only when the opinions can take more 
values than just $\pm 1$ (e.g., $\pm 1$ and $0$). 
In our opinion dynamics model, the ordered phase may be  regarded as
a ferromagnetic phase and the disordered phase  as a paramagnetic phase
(certainly not a spin glass phase as the opinions
do not attain a frozen state had it been so).

When agents are `selective' in their interactions and
some of the interactions are absent or muted, 
this effectively makes those $\mu_{ij} =0$. 
If $\mu_{ij} =0$ with probability $q$, then we can conceive a phase diagram with respect to 
the parameters $p$ and $q$ (see Fig.~\ref{fig:dilute}) separating the ordered and disordered 
phases.

A comparison of our  
model with the  
model  of Ref \cite{galam1} which introduces Galam contrarians (see also \cite{wio,stau-martins,galam2}) may also be made.
In the latter, one has two discrete choices of opinion 
while the general case of the present model involves  
continuous opinion values. The mean-field critical behavior, however, is observed in both models 
with same exponent values.
In Ref \cite{galam1} the contrarians  
would take the 
opinion which is exactly opposite of that of  the majority of the group. The evolution of opinions  is clearly 
different in our case, where the original opinion of an   agent 
is also considered while assigning the changed value. This makes even the discrete
version of the present model  consist of three states (-1, 0 +1). Furthermore, we have shown that (see Eq. (\ref{diso}))
if we consider the discrete version to have two states only, then there is no ordered state in the model as soon as any
finite fraction of negative interactions is introduced.  

Results obtained from opinion dynamics models can be compared to real data to a certain extent. Of course, the microscopic rules governing the dynamics in a model cannot be verified directly but one can justify the model by comparing its macroscopic behavior with real data. Discrete opinion models may be applicable to election scenarios \cite{galam1}.
Continuous opinion models, on the other hand, mimic the case of rating or degree of support for an issue. So in our model, -1 represents extreme unfavorable opinion, +1 is extreme favorable while zero means an average rating/indifferent response.
Thus the order parameter in the model  corresponds to the overall rating and ordered state means there is a clear-cut decision made. A disordered state means the absence of a decision. The situation is comparable to a case of final verdict arrived at by a panel of juries. If there is a lot of disagreement among the
juries  a decision is hard to achieve - this result is
indeed obtained in our model where the disagreement is
represented by the negative interaction terms.

Although we have considered continuous opinions in the model, it was
shown exactly that using discrete opinions 1, -1, 0 also leads to the same critical behavior.
From the statistical physics viewpoint, the important role in the phase transition
is thus played by the parameter $p$ quantifying the fraction of negative interactions, irrespective of the fact
that opinions are continuous or discrete. This precisely indicates that $p$ is the relevant
parameter in the model and the nature of opinion is irrelevant as far as critical behavior is concerned

We conclude with the remark that the values of the exponents $\beta, \gamma$ 
are very similar to that of the mean field exponents of the Ising model. 
Interpreting $\nu $ as  $\nu^{\prime} d$ where $d$ is the effective dimension  
in this long ranged model and putting $d=4$  as is done in small world like 
networks \cite{small-world}, the value of the effective correlation length 
exponent $\nu^\prime = 1/2$ also coincides with the mean field value.

For future study, the properties of this model in various lattices and networks
and also its dynamical behavior would be interesting   \cite{work}.

The authors would like to specially thank H. Nishimori for very useful
correspondence.  
Inspiring discussions with B. K. Chakrabarti, J. Kert\'{e}sz and M. Marsili
are also acknowledged. 
SB acknowledges discussions with N. Sarkar.
PS acknowledges  UPE project and DST project for 
partial support.

\bibliographystyle{elsarticle-num}
\bibliography{<your-bib-database>}

\end{document}